\documentclass[pdflatex,sn-mathphys-num]{sn-jnl}

\usepackage{graphicx}%
\usepackage{multirow}%
\usepackage{amsmath,amssymb,amsfonts}%
\usepackage{amsthm}%
\usepackage{mathrsfs}%
\usepackage{xcolor}%
\usepackage{ulem}%
\usepackage{textcomp}%
\usepackage{manyfoot}%
\usepackage{booktabs}%
\usepackage{listings}%
\usepackage{physics}%
\usepackage{bm}%
\usepackage{float}%

\begin{document}

\title[Brazilian Twin Photons 32nd anniversary]{Brazilian Twin Photons 32nd anniversary}


\author*[1]{\fnm{Renné} \sur{Medeiros de Araújo}}\email{renne.araujo@ufsc.br}

\author[1]{\fnm{Raphael César} \sur{Souza Pimenta}}
\equalcont{These authors contributed equally to this work.}

\author[1]{\fnm{Lucas} \sur{Marques Fagundes}}
\equalcont{These authors contributed equally to this work.}

\author[2,3]{\fnm{Gustavo Henrique} \sur{dos Santos}}
\equalcont{These authors contributed equally to this work.}

\author[1]{\fnm{Nara} \sur{Rubiano da Silva}}
\equalcont{These authors contributed equally to this work.}

\author[2,3]{\fnm{Stephen Patrick} \sur{Walborn}}
\equalcont{These authors contributed equally to this work.}

\author[1]{\fnm{Paulo Henrique} \sur{Souto Ribeiro}}
\equalcont{These authors contributed equally to this work.}

\affil*[1]{\orgdiv{Departamento de Física}, \orgname{Universidade Federal de Santa Catarina}, \orgaddress{\city{Florianópolis}, \postcode{88040-900}, \state{SC}, \country{Brazil}}}

\affil[2]{\orgdiv{Departamento de Física}, \orgname{Universidad de Concepción}, \orgaddress{\city{Concepción}, \postcode{160-C}, \state{Bío Bío}, \country{Chile}}}

\affil[3]{\orgdiv{Millennium Institute for Research in Optics}, \orgname{Universidad de Concepción}, \orgaddress{\city{Concepción}, \postcode{160-C}, \state{Bío Bío}, \country{Chile}}}


\abstract{We present a historical review of the development and impact of spontaneous parametric down-conversion (SPDC) in Brazil, marking over three decades since the first twin-photon experiments were performed in the country. This article traces the pioneering efforts that initiated the field, highlighting key experiments, institutions, and researchers who contributed to its growth. We discuss seminal works that established SPDC as a fundamental tool in the Brazilian Quantum Optics community, including studies on spatial correlations, entanglement, and decoherence. By presenting a curated sequence of experiments, we offer an overview of how Brazilian research in twin-photon systems has explored profound concepts through fundamental demonstrations, leading to significant international impact. This review also highlights the formation of a strong scientific community and its ongoing efforts to turn fundamental knowledge into quantum applications.}




\maketitle

\section{Introduction}
Spontaneous parametric down-conversion (SPDC) is a nonlinear process where one photon from an intense pump beam is converted into two photons of lower frequencies, such that energy and momentum are conserved, giving rise to correlations between the properties of the individual photons. The process and the first applications in the field of Quantum Optics were predicted by the Russian physicist D. N. Klyshko \cite{zeldovich69}, while the first experimental investigation observing the notable correlation properties of the photon pairs was performed by the North American physicists Burnham and Weinberg in 1970 \cite{Burnham70}. They observed that the pair of photons emitted in SPDC was always detected simultaneously, and concluded therefore they were ``born" simultaneously in the medium.  This originated the term ``twin photons".

The German-American physicist Leonard Mandel performed a series of SPDC twin photon experiments in the 1980s and discovered many important quantum effects. The most popular of these quantum effects is the so-called Hong-Ou-Mandel effect \cite{hom87}, which, since its discovery, has been widely used in several applications, including teleportation and entanglement swapping. Moreover, it was shown that one can use SPDC to generate polarization-entangled states, also known as Bell states \cite{Ou88}. The use of these twin photons in experiments studying important foundational concepts of Quantum Mechanics strongly contributed to the Nobel prize awarded to the Austrian physicist Anton Zeilinger in 2022. The prize was shared with John Clauser, who in the 1970s performed experiments with twin photons, and with Alain Aspect, who carried out similar experiments in the early 1980s. In both cases, instead of SPDC they used cascaded atomic emission, which is much less reliable and efficient than SPDC.

In Brazil, SPDC was introduced by Geraldo A. Barbosa at the Federal University of Minas Gerais in the early 1990s. The first experiments using twin photons from SPDC and coincidence detection schemes at UFMG were published by Carlos H. Monken and G. A. Barbosa \cite{monken93} in 1993 and P. H. Souto Ribeiro, S. Pádua, J. C. Machado da Silva, and G. A. Barbosa in 1994 \cite{ph94}.  These first contributions were instrumental in the establishment of a robust experimental platform and in the development of a broad and influential body of research. One important achievement was the theoretical derivation and experimental verification of the {\it Transfer of the Angular Spectrum} from the pump beam to the quantum correlations shared by the twin photons, by C. H. Monken, P. H. Souto Ribeiro and S. Pádua \cite{monken98}.  Its significance is reflected in the breadth of work it has inspired, both in Brazil and internationally. 

In this International Year of Quantum Science and Technology (IYQ), we found it appropriate to highlight the 32 years of experiments with twin photons in Brazil. This field can be considered a scientific success story, as it has involved over 100 Brazilian researchers, produced roughly 100 graduate dissertations and theses, and led to 300+ publications, many in the top physics and optics journals.  We attribute these prolific results to several factors.  First, Brazil experienced more than a decade of stable scientific funding starting in the early 2000s, including Millennium Institutes and INCTs focused on quantum optics. Second, the development of the field of quantum information led to increased interest in new phenomena involving quantum optical coherence, such as entanglement and quantum correlations, as well as interesting applications. Lastly, we believe that ``Brazilian ingenuity'' played a key role in finding a way to do interesting and competitive science with fewer resources than other international competitors in the field.

In the following sections, we describe selected experiments with the {\it Brazilian Twin Photons}, emphasizing the novel results achieved, the establishment of new laboratories in Brazil, and the formation of PhD students who are today leading research in Brazil and abroad.

\section{Building the first twin photon laboratory}
The twin-photon coincidence counting technique was first introduced at UFMG by Prof. G. A. Barbosa, who, together with his PhD student C. H. Monken, constructed an experimental setup capable of generating and detecting photon pairs in coincidence. This pioneering effort led to the first experiment of its kind in Brazil, and quite possibly the first outside the United States and Europe.

In their setup, signal and idler photons were produced via spontaneous parametric down-conversion in a bulk lithium iodate (\(\text{LiIO}_3\)) crystal, pumped by an argon-ion laser operating at 351 nm. The down-converted photons -- referred to as \textit{signal} and \textit{idler} -- were detected using photomultiplier tubes, while the coincidence measurements were performed using analog electronics typical of high-energy physics experiments.

\subsection{Photon ricochet in a cavity}

\begin{figure}[h!]
	\centering 	\includegraphics[width=0.7\columnwidth]{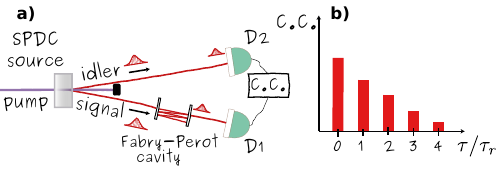}
	\caption{Photon ricochet in a cavity: measurement of the photon time of flight in a non-interfering linear cavity \cite{monken93}. a) Conceptual representation of the  experimental setup and b) results.}
	\label{ricochet}
\end{figure}

In the first experiment, they used the temporal correlation between the photons to investigate the transit of a single photon in an optical cavity.
The idler photon was directed through an optical cavity in the non-interfering regime, while the signal photon was sent directly to the detector. The experimental setup is illustrated in Fig. \ref{ricochet}a). The time difference between signal and idler detections was recorded using a time-to-amplitude converter (TAC). The key result was a histogram similar to the one shown in Fig. \ref{ricochet}b), featuring peaks with decaying amplitudes as the time difference increased. Each peak corresponded to a photon that underwent a complete round trip $\tau_r$ within the cavity. Since the experiment was conducted in the non-interfering regime, the photonic wavepacket behaved like a particle undergoing successive reflections between the cavity mirrors. The decreasing peak amplitudes were attributed to mirror losses upon each reflection. The findings were published in Optics Communications \cite{monken93}.

\section{Spatial Correlations and Coherence}

Most of the SPDC experiments in the 1980s, particularly by L. Mandel's group, focused on the temporal or frequency correlations of the SPDC photon pairs.  In the early 1990s, a few research groups turned their attention to the spatial properties.  In Brazil, several seminal results were produced during the 1990s and early 2000s. Novel and curious properties concerning the spatial degrees of freedom of the photon pairs were studied, including the appearance of a de Broglie wavelength of a two-photon wavepacket, and quantum images that appear only in the coincidence count distributions. Many of these studies occurred before the theory of entanglement was developed.  In many cases, the quantum nature of the observations was demonstrated through the conditional character of the coincidence count distribution, which depends on the positions of both detectors. Overall, these studies contributed to the development of the \textit{biphoton} concept -- the idea that a two-photon wavepacket exhibits behavior characteristic of a single quantum entity.

\subsection{Nonlocal control of spatial coherence}

\begin{figure}[h!]
    \centering\centering 	\includegraphics[width=0.7\columnwidth]{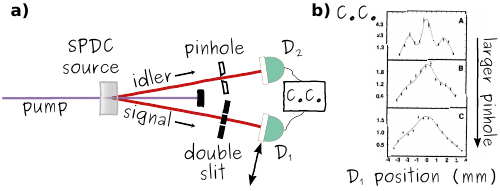}
    \caption{Nonlocal control of spatial coherence: a) Conceptual representation of the experimental setup and b) results extracted from article \cite{ph94}. The coincidence counts show the interference pattern for different diameters of pinhole. }
    \label{fig_conhence_control}
\end{figure}

As part of his doctoral thesis, P. H. Souto Ribeiro, under supervision of Geraldo A. Barbosa, and collaboration of S. P\'adua and J. C. Machado da Silva, demonstrated that the visibility of the interference fringes in a Young double slit experiment could be remotely controlled by using twin photons and coincidence counting \cite{ph94}. Figure \ref{fig_conhence_control}a) presents the experimental setup used: signal and idler photons are produced in SPDC. The idler photon is sent through a double slit, while the signal photon passes through a pinhole of variable diameter before detection. A typical double slit interference pattern is obtained by displacing the idler detector while keeping the signal one fixed, and registering the coincidence counting rate. The interesting result is that the visibility of the interference patterns depends on the diameter of the pinhole placed in front of the signal detector, whose position is kept fixed. Therefore, the coherence of the idler photon is remotely controlled by the way the signal photon is detected. This property stems from the spatial correlations between signal and idler photons so that the signal pinhole acts as a filter for the angular spectrum of the idler, allowing the control of the visibility of the interference fringes. A smaller pinhole leads to larger visibility, as illustrated in Figure \ref{fig_conhence_control}b).

 The scientific contribution of the article lies in the use of a coincidence detection scheme to observe interference fringes with high contrast, even when the transmitted light has a transverse coherence length that is much smaller than the slit separation. 
 A notable extension of this study was performed in 2012 by Almeida \textit{et al.}, who showed that when the pump beam is prepared as a Hermite-Gaussian mode, in some cases, a large pinhole can actually lead to a larger visibility \cite{almeida12}. This effect is due to a spatial mode parity selection that depends on the pinhole diameter \cite{walborn11b}.

\subsection{Transfer of the angular spectrum}
\label{sec:transf-ang-spec}

In the late 1990s, Carlos H. Monken developed a theory describing the quantum state of the transverse spatial degrees of freedom of photon pairs, showing the transfer of the angular spectrum of the pump beam to the quantum state of the signal and idler photons. To demonstrate this remarkable and far-reaching result, an experiment was carried out in 1998 at UFMG under the leadership of Carlos H. Monken, in collaboration with Sebastião Pádua and Paulo H. Souto Ribeiro (then a postdoctoral researcher).

\begin{figure}[h!]
	\centering 	\includegraphics[width=0.7\columnwidth]{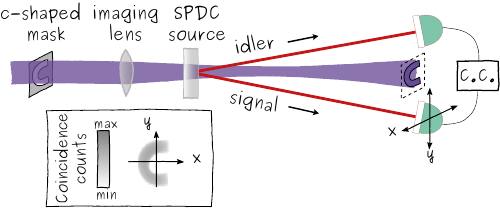}
	\caption{Transfer of the angular spectrum: conceptual representation of the experimental setup and results. Measurement of the transfer of the angular spectrum from the pump to the quantum correlations \cite{monken98}. }
	\label{transg}
\end{figure}

The experiment is illustrated in Fig. \ref{transg}. The pump laser was sent through a binary mask with a C-shaped aperture,  and then to a non-linear crystal. Signal and idler photons are detected in coincidence.  A lens was used to form an image of the pump beam at the same distance from the crystal as the detectors.  Then, with one detector kept fixed and the other scanned across the transverse detection plane, the coincidence distribution was observed to reproduce the shape of the letter “C” encoded in the pump. In other words, the angular spectrum of the pump was “imprinted” onto the transverse spatial (conditional) correlations between the twin photons. This study stands as one of the most important in the field and has served as the foundation for numerous works carried out both in Brazil and abroad.

The angular spectrum transfer provides a way to engineer quantum states with different interesting properties, some of which will be described in the following sections. A first application of these results was a method to increase the collection efficiency of the photon pairs by focusing the pump beam in the detection plane, thus tightening the spatial correlation and increasing the coincidence counts \cite{monken98b}.  What is miraculous is, as a colleague communicated to one of us a decade later\footnote{Dr. Andrew White, private communication to SPW.}, ``the single photon counts stay the same!", such that the detection efficiency ($\sim$ coincidences/singles) increases. Other applications include predicting spatial parity correlations \cite{walborn05}, controlling spatial correlations with polarization \cite{Caetano2002,Caetano2003}, and increasing the spatial entanglement of the two-photon state by using a structured pump beam \cite{walborn07pra}.   

Building upon the original theory, Brazilian researchers extended these results by applying them to stimulated down-conversion \cite{ribeiro99}, including the effects of anisotropy of the non-linear crystal \cite{Moura2010}, considering 
angular spectrum transfer for vector pump beams \cite{khoury2020}, and also for partially coherent pump beams \cite{hutter2020,hutter2021}.

\subsection{Nonlocal double slit}

The nonlocal double-slit experiment \cite{fonseca99} was also carried out at UFMG, based on the theory of angular spectrum transfer. At the time, Eduardo J. S. Fonseca was a PhD student under the supervision of Sebastião Pádua and, together with Paulo H. Souto Ribeiro (then a postdoctoral researcher) and Carlos H. Monken, performed a double-slit interference experiment in which one part of the slit was placed in the signal path and the other in the idler path. In this setup, the signal and idler photons were each scattered by diffractive objects that individually did not constitute a double slit. However, when combined nonlocally, they produced a double-slit interference pattern observable in the coincidence counting rate, even though no interference fringes appeared in the local intensity distributions.

\begin{figure}[h!]
	\centering 	\includegraphics[width=0.7\columnwidth]{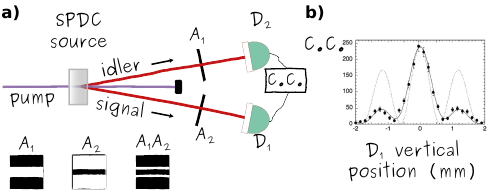}
	\caption{Nonlocal double slit: a) Conceptual representation of the  experimental setup and b) the results extracted from Ref. \cite{fonseca99}. }
	\label{fig:nonlocaldoubleslit}
\end{figure}

Figure \ref{fig:nonlocaldoubleslit}a) shows the experimental setup used. The apertures $A_1$ and $A_2$ are placed in the paths of the idler and signal photons, respectively. $A_1$ is a single slit, and $A_2$ is an opaque rectangle. The product of these two amplitude transmission functions $A_1A_2$ corresponds to a double slit diffraction object.  Figure \ref{fig:nonlocaldoubleslit}b) illustrates the expected behavior of the coincidence counting rate, when detector $D_1$ is displaced and $D_2$ is kept fixed. The contribution of this work lies in fact that the principle of nonlocality is explicitly demonstrated through the interference pattern observed in the coincidence profile that can be obtained by displacing anyone of the detectors and keeping the other fixed.

\subsection{Multiphoton de Broglie wavelength}

Twin photons are often referred to as a \textit{biphoton}, since their correlations are so strong that their properties are only well defined when considered jointly, allowing them to be treated as a ``single entity'' rather than as two independent photons. A striking demonstration of this behavior is the emergence of an effective de Broglie wavelength equal to approximately half that of each individual photon\footnote{The concept of the de Broglie wavelength, originally formulated for atoms and molecules, was generalized to an $n$-photon wave packet by Yamamoto \cite{jacobson95}.}.

\begin{figure}[h!]
	\centering 	\includegraphics[width=0.7\columnwidth]{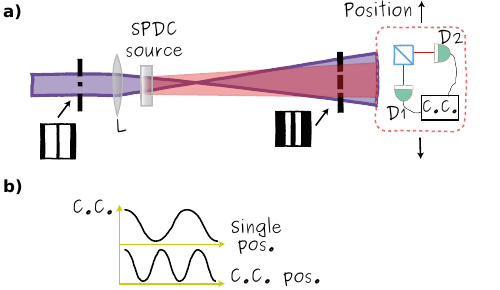}
	\caption{Multiphoton de Broglie wavelength: conceptual representation of the a) experimental setup and results, and b) measurement of the DeBroglie wavelength of a biphoton \cite{fonseca99b}. }
	\label{debrog}
\end{figure}

To demonstrate this effect experimentally, Eduardo J. S. Fonseca, Sebastião Pádua, and Carlos H. Monken performed the experiment shown in Fig.~\ref{debrog} \cite{fonseca99b}. The setup consists of a double-slit interference arrangement in which twin photons produced via SPDC are prepared to behave as a two-photon wavepacket. The suggestion to relate the de Broglie wavelength to a two-photon double-slit experiment at UFMG was made by Luis Davidovich to Sebastião Pádua during a visit to UFMG in 1998. This condition is achieved by tailoring the angular spectrum of the pump beam such that the spatial correlations between signal and idler photons force them to propagate together through the slits, suppressing events in which the photons are separated. As a consequence, the interference fringes oscillate with a spatial frequency determined by the de Broglie wavelength of the biphoton wavepacket, rather than by the single-photon wavelength. This behavior was demonstrated by measuring the interference fringes under two conditions: first, with a standard Gaussian pump beam, as illustrated in the upper plot of Fig. \ref{debrog}b), and second, with a pump beam engineered with the appropriate angular spectrum. The latter case clearly reveals the higher oscillation frequency of the fringes, as shown in the lower plot of Fig. \ref{debrog}b).

In 2001, Eduardo Fonseca, Zoltan Paulinyi (then a master’s student supervised by S. Pádua), Paulo Nussenzveig, C. H. Monken, and S. Pádua extended the concept of the de Broglie wavelength to composite systems whose constituent particles are spatially separated, and demonstrated it for a two-photon system \cite{fonseca2001}.

\subsection{Which-crystal interference}

In 2001, P. H. Souto Ribeiro published the results of the first experiment carried out at the newly established Quantum Optics Laboratory at the Physics Institute of the Federal University of Rio de Janeiro \cite{ph01}. The experiment, conducted by Souto Ribeiro at UFRJ with equipment partially borrowed from colleagues at UFMG, demonstrated two-photon interference without the use of slits or beam splitters. Despite the limited resources available, the work combined conceptual clarity with an innovative design, marking an important milestone for the laboratory.

\begin{figure}[h!]
	\centering 	
        \includegraphics[width=0.7\columnwidth]{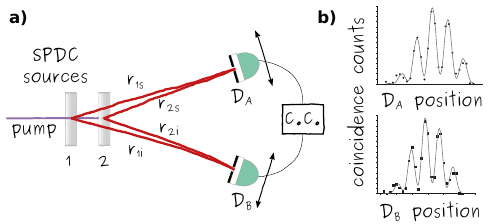}
	\caption{Which-crystal interference: a) experimental setup for observing which-crystal interference. b) Coincidences as a function of the detectors displacement \cite{ph01}.}
	\label{fig:which-crystal}
\end{figure}

The setup employed two crystals pumped by a continuous-wave laser to produce photon pairs via SPDC (see Fig.~\ref{fig:which-crystal}). Interference fringes emerged purely from spatial correlations in the biphoton wave function, as the relative positions of the detectors placed after each crystal were varied.

A key insight of the study was that the effective wavelength associated with the interference -- the ``biphoton wavelength” -- is not fixed, but depends on the detection strategy. By adjusting the way detectors are scanned, the resulting wavevector can correspond to single-photon interference, two-photon interference, or even intermediate (fractional) values. This behavior does not result from any energy shift, but rather from entanglement and quantum correlations, offering a compelling platform for studying fundamental aspects of quantum measurement.

This early demonstration inspired further investigations into spatial quantum correlations without relying on traditional interferometric elements. Building on this approach, several follow-up studies explored the control and manipulation of spatial entanglement. Noteworthy examples include ``Quantum distillation of position entanglement with the polarization degrees of freedom” \cite{Caetano2002} and ``Image formation by manipulation of the entangled angular spectrum” \cite{caetano2004}, both published in Optics Communications, as well as ``Entanglement of the transverse degrees of freedom of the photon”, published in the Journal of Optics B \cite{souto_ribeiro2002}.

\subsection{Double slit quantum eraser}

\begin{figure}[h!]
	\centering 	\includegraphics[width=0.7\columnwidth]{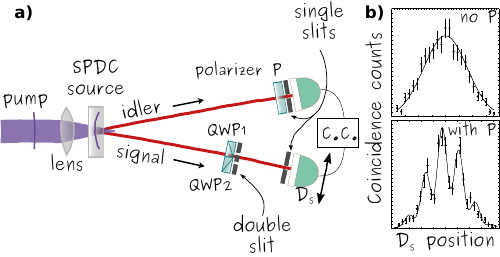}
	\caption{Double slit quantum eraser: a) conceptual representation of experiment to observe the quantum eraser. b) Results showing the recover of double slit pattern when a polarizer is placed.}
	\label{expqe}
\end{figure}

Quantum erasure is an intriguing and paradigmatic phenomenon. It highlights the informational character of Quantum Theory and the importance of the measurement. During the PhD thesis of S. P. Walborn at UFMG under the supervision of C. H. Monken, they conceived and realized an experiment in cooperation with Marcelo Terra Cunha (who had originally brought the idea to C. H. Monken) and S. Pádua. They observed the effect of quantum erasure in a double slit experiment using twin photons prepared in polarization-entangled Bell states. The experimental setup is illustrated in Fig. \ref{expqe}.  

\begin{figure}[h!]
	\centering 	\includegraphics[width=6cm]{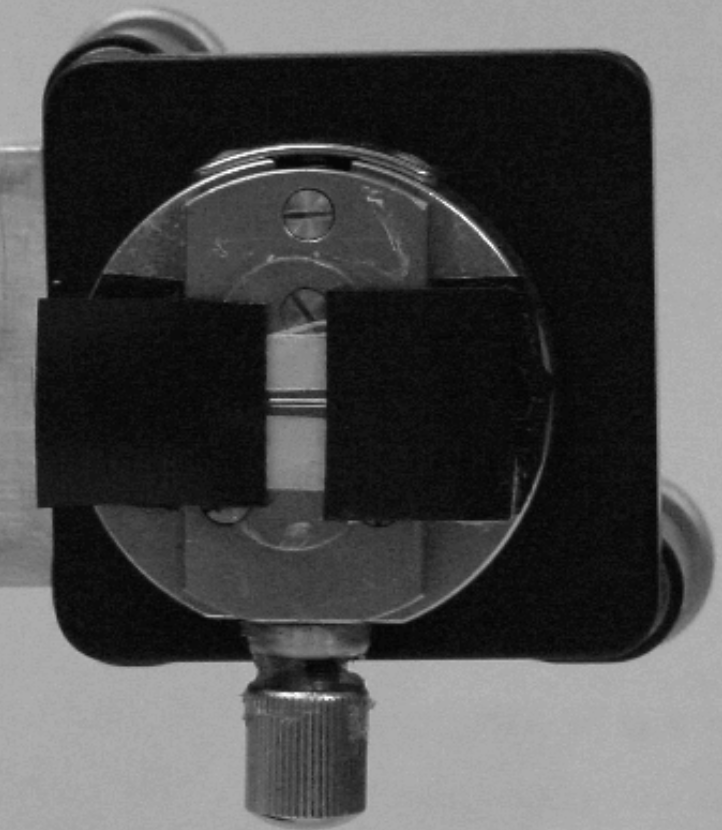}
	\caption{The birefringent double slit was fabricated in 1999 and used in several experiments \cite{walborn02,nogueira01,nogueira02,Neves2009,Torres2010}. The slits were 0.2~mm wide, constructed by placing a 0.2~mm wire at the center of a 0.6~mm slit salvaged from a defunct device. To accommodate circular wave plates in front of the slits, one of us (SPW) sanded a flat chord along the edge of the plates. For this purpose, the wave plates were first orthogonally aligned using polarizers and a laser, and then attached together with double-sided tape. To ensure that the sanding produced a straight edge rather than a curved one, C. H. Monken devised a simple apparatus by removing the springs from an optical translation stage, allowing the plates (fixed to the stage with double-sided tape) to be moved freely back and forth across diamond polishing paste. A small weight was added to provide the proper pressure. In this way, SPW manually sanded the plates for about 10 hours until a straight chord, approximately 5 mm long, was obtained. In the first attempt, however, the plates were misaligned for use in the quantum eraser experiment \cite{walborn02}, but -- by good fortune -- this alignment turned out to be the correct configuration for the spatial antibunching experiment \cite{nogueira01}.}
	\label{fig:bds}
\end{figure}

The key element of this experiment was a specially designed double slit. It was implemented by placing two quartz waveplates in front of a Young's double slit, with each waveplate marking the photon's path by rotating its polarization. A photograph of the device, along with a detailed description, is provided in Fig. \ref{fig:bds}.

In this experiment, twin photons were produced in SPDC using a BBO crystal cut for type II interaction. Using this scheme it is possible to prepare Bell states for the polarization state like $\ket{\Phi^+} = \frac{1}{\sqrt{2}} \left( \ket{00} + \ket{11} \right)$, where $\ket{0}$ and $\ket{1}$ represent two pure and orthogonal polarization states. One of the photons, say signal, was sent to the special double slit with QWP1 and QWP2 waveplates, and the other one, the idler, to detection, after crossing a linear polarizer P. The signal detector is scanned in order to measure the interference pattern by registering coincidences between signal and idler photon counts. When P is removed, no interference fringes are observed in the coincidence pattern. This is because the plates QWP1 and QWP2 change the polarization state at each slit and therefore provide which-path information. When the plates are removed, the double slit interference pattern shows up, because the which-path information is no longer available. However, if the QWP1 and QWP2 are in place and P is used in front of the idler detector, the which-path information can be remotely erased. For a given orientation of P the interference fringes are recovered. For the orthogonal orientation they also show up with a dephasing of $\pi$ and for a certain orientation the fringes disapear again.  

In conclusion, the which-path information can be remotely controlled by means of the entanglement between signal and idler photons and proper polarization state measurement of the idler photon. The results were published in Physical Review A \cite{walborn02}.

\subsection{Spatial anti-bunching of photons}

The experiment reported in Ref. \cite{nogueira01} was published before the double-slit quantum eraser experiment \cite{walborn02}, although it was actually performed later (see caption of Fig. \ref{fig:bds}). It employed the same birefringent double-slit aperture shown in Fig. \ref{fig:bds}. Spatial photon antibunching was subsequently observed in a similar configuration with twin photons, but in free propagation without the use of double slits \cite{caetano03}.

Photon antibunching in stationary fields provides an unambiguous signature of nonclassicality, since such states cannot be described by a positive Glauber–Sudarshan $P$ distribution. In contrast, stationary states exhibiting bunching statistics and random behavior admit a classical representation through this distribution. The work in Ref. \cite{nogueira01} advanced the field by establishing a spatial regime in which the violation of classical inequalities offers clear evidence of spatial photon antibunching.

\begin{figure}[H]
\centering 	\includegraphics[width=0.7\columnwidth]{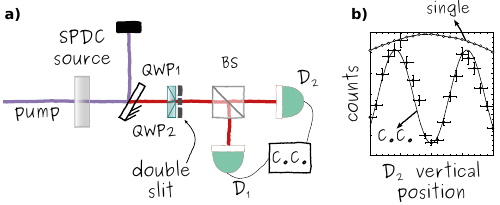}
	\caption{Spatial anti-bunching of photons: a) conceptual representation of experiment to observe spatial anti-bunching of photons. b) The main results show that, in coincidence counts, the Schwarz inequality is violated.}
	\label{exp:antibuching}
\end{figure}

The paper presents experimental evidence of spatial antibunching along one direction in the transverse plane relative to the propagation axis, showing a clear violation of the Cauchy–Bunyakovsky–Schwarz inequality. The effect was observed in the fourth-order correlation function of down-converted photons diffracted by a double slit (see Fig.~\ref{exp:antibuching}). The measured correlation function -- proportional to the coincidence rate between two point detectors (under appropriate approximations) -- reveals the occurrence of antibunching. By analyzing both single counts and coincidence rates, the authors demonstrated systematic violation of the classical inequality, as shown in Fig. \ref{exp:antibuching}b).

\subsection{Multimode fourth-order interference}

\begin{figure}[h!]
	\centering 	\includegraphics[width=0.7\columnwidth]{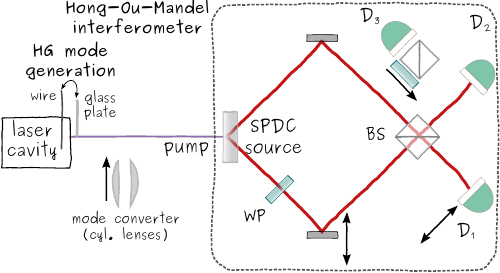}
	\caption{Multimode fourth-order interference: conceptual representation of the experimental setups used in  \cite{walborn04,nogueira04,walborn03}. A HG mode laser was generated employing either a wire \cite{walborn04,walborn03} or a glass plate \cite{nogueira04}, placed inside or just after the laser cavity, respectively. A couple of cylindrical lenses could be introduced to convert the beam to a LG mode before pumping the SPDC process. A Hong-Ou-Mandel interferometer with variable path length difference was used to observe fourth-order correlations between detectors D$_1$, D$_2$ and/or D$_3$ (if inserted). A half \cite{walborn04,walborn03} or a quarter \cite{nogueira04} waveplate (WP) was introduced in one of the interferometer arms.}
	\label{fig:multimode}
\end{figure}

In Ref.~\cite{monken98} (see section~\ref{sec:transf-ang-spec}), it has been demonstrated that the angular spectrum of the pump beam is transferred to the two-photon state, and the intensity profile of the pump is transferred to fourth-order spatial correlations of twin photons at the detector planes for the degenerate frequency (i.e. the spatial distribution of the coincidence counts). Among the many works that followed, there are observations of different effects when taking into account the spatial modes, either pure or superposed, of the signal and idler photons. Figure~\ref{fig:multimode} presents the conceptual illustration of the experiments that we describe in the following subsections.

\subsubsection{Entanglement and conservation of OAM in SPDC}
\label{sec:oam}

In a study involving theoretical and experimental demonstrations, Steve P. Walborn, Alvaro Nunes de Oliveira and Rafael Santos Thebaldi, graduate students at UFMG at that time under the supervision of Carlos H. Monken, showed that orbital angular momentum (OAM) is not only conserved in SPDC, but also leads to a biphoton state with OAM entanglement \cite{walborn04}.

The experiment consisted in two basic parts (see Fig.~\ref{fig:multimode}): the generation of a Laguerre-Gaussian (LG) beam and a Hong-Ou-Mandel interferometer. To create the LG pump beam, a 25~$\mu$m gold wire was placed inside the laser cavity to break the cylindrical symmetry.  Depending on the position of the wire, low-order Hermite-Gaussian (HG) modes could be produced.  Using a lens-based mode converter \cite{beijersbergen93}, the HG modes were converted to LG modes. The beam then pumped a BBO crystal producing signal and idler photons, whose correlations were measured using the interferometer. The coincidence patterns were compared to theoretical expectations, and observed to show spatial correlations only with a balanced interferometer, thus demonstrating the entanglement of OAM.

\subsubsection{Multimode Hong-Ou-Mandel}
\label{sec:MMHOM}

When performing the experiment in section \ref{sec:oam}, some of the authors realized that the two-photon images in the Hong-Ou-Mandel interference were indeed the Hermite-Gauss components of the Laguerre-Gauss two-photon transverse profile. By scanning the path length difference of the interferometer arms in Fig.~\ref{fig:multimode} for different combinations of pump modes and photon pair polarization, Steve Walborn, Alvaro Oliveira, S. Pádua and Carlos H. Monken demonstrated the relation between the parity of the pump beam and the spatial symmetry of the two-photon state, which governed two-photon Hong-Ou-Mandel interference \cite{hom87}. This led to a study of this effect \cite{walborn03}. An application of these results is the determination of three classes of polarization Bell-states in the coincidence basis \cite{Walborn03a}.

\subsubsection{Two-photon singlet beam}

Based on the results obtained in Ref. \cite{walborn03}, Wallon Nogueira, Steve Walborn, Sebastião Pádua and Carlos H. Monken observed that  photons produced with an anti-symmetric polarization state (singlet) and an anti-symmetric spatial state would interfere destructively at the BS, and exit together in the same exit port. Specifically, the observation was performed by generating photon pairs via SPDC in two polarization Bell states, and using an additional detection scheme (HWP, PBS and $D_3$ in Fig.~\ref{fig:multimode}) to identify the beam splitter output port of the photons. The quantum state in each output thus constituted a ``singlet beam", which presents several interesting properties \cite{nogueira04}. First, the beam is completely unpolarized, such that the first-order moment of the Stokes vector is zero. However, the beam presents higher-order polarization, as the photon pair is in a well-defined polarization state.  A second effect is that the spatial asymmetry naturally leads to spatial anti-bunching, as observed in Ref. \cite{nogueira01}.  We note that spatial antibunching in a similar beam-like setup was also produced in Ref. \cite{caetano03}.   
\par
From a technical standpoint, this experiment was the first to use a thin microscope slide to produce a $\pi$ phase shift between two halves of the pump beam, thus avoiding the need to modify the laser cavity (as in sections \ref{sec:MMHOM} and \ref{sec:oam}; see Fig.~\ref{fig:multimode}), leading to a simpler and more robust technique for producing the asymmetric spatial state of the photon pair. This was used in the experiments discussed in section \ref{sec:beyond}, as well the observation of anomalous coherence effects from nonlocal spatial filters \cite{almeida12}. 

\section{Spatial Entanglement}

Around 2000, the theory of entanglement and, in particular, continuous-variable entanglement was in development.  As a result, a number of practical entanglement criteria \cite{reid89,duan00,mancini02}  appeared and were applied to the spatial correlations of photon pairs. This added new understanding to the previous work on spatial correlations and coherence, leading to the discovery of new phenomena, and making links with quantum information concepts and applications. 
Entanglement appeared via the fact that the photons were simultaneously correlated in the near-field of the crystal (the photons are ``born" together) and anti-correlated in the far-field (due to momentum conservation), as demonstrated in Ref. \cite{howell04}.

\subsection{Entangled qudits}

\begin{figure}[h!]
	\centering 	\includegraphics[width=0.7\columnwidth]{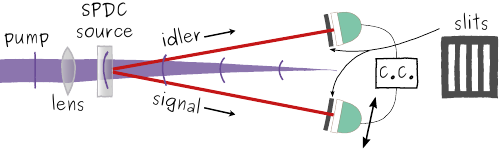}
	\caption{Entangled qudits: experimental setup used to explore spatial correlations to produce qudits with multiple slits and a focusing pump beam.}
	\label{fig:qudits-sebastiao}
\end{figure}

In 2005, Leonardo Neves and Gustavo Lima were PhD students under the supervision of Sebastião Pádua at UFMG and together with Carlos H. Monken (UFMG), Aguirre Gomez, postdoc at UFMG, and Carlos Saavedra from the University of Concepción, Chile, they realized an experiment that represented a significant advance in the manipulation of high-dimensional quantum systems using spatial degrees of freedom of photons generated via SPDC. In “Generation of entangled states of qudits using twin photons” \cite{neves05}, the authors implemented a novel method to generate maximally entangled states of qudits -- quantum systems with dimension $D>2$ -- by inserting multi-slit apertures into the paths of each photon, close to the detectors, as illustrated in Fig.~\ref{fig:qudits-sebastiao}. These slits define discrete spatial modes that represent the computational basis of the qudits. The core idea of the experiment is to engineer the spatial correlations such that the twin photons emerge only through symmetrically opposite slits, thereby producing path-entangled qudit states of the form $\sum_{l=-l_D}^{l_D} |l\rangle_s|-l\rangle_i$, where $s$ and $i$ denote signal and idler photons and $|l\rangle$ represents a single photon through the slit $l$ and $l_D=(D-1)/2$.

A key ingredient in this spatial entanglement control is the careful shaping of the pump beam, which determines the spatial correlations of the down-converted photons through momentum conservation. In particular, by tightly focusing the pump beam at the plane of the multi-slit apertures, the authors ensured that the crystal acted like a concave ``quantum mirror'' allowing only certain spatial correlations to survive. This phenomenon can be intuitively understood using Klyshko’s advanced-wave picture \cite{belinskii94}: if one imagines an advanced wave propagating backward from one detector, it would be reflected by the nonlinear crystal as if the crystal were a concave mirror, focusing the wave toward the other detector. This optical analogy elegantly explains how the spatial mode of one photon determines that of its twin, enabling the selective generation of entangled states across symmetric slits.

The experiment demonstrated entanglement for qudit dimensions $D=4$ and $D=8$, providing evidence of the spatial coherence and quantum nature of the twin-photon states. The ability to generate and control entangled qudit states is crucial for enhancing the information-carrying capacity of quantum systems and for improving the robustness of quantum communication schemes against noise. At the time, few experimental platforms were capable of generating such high-dimensional entanglement in a compact and controllable way, making this work particularly impactful.

This contribution from Sebastião Pádua's group laid the groundwork for a series of subsequent studies, including investigations of photonic qudit propagation \cite{lima06}, generation of mixed states via pump beam preparation \cite{lima08a}, state tomography of qubits and qutrits \cite{neves07,lima08b,pimenta13}, and entanglement detection using modular variables \cite{carvalho12}. The latter was, to the best of our knowledge, the first experiment to employ full-field interference patterns as a witness of entanglement.

More recently, Pádua and collaborators replaced the slits in front of the detectors with a multi-path pump beam -- implemented using beam-displacing prisms -- to generate similar quantum states and reproduce the characteristic qudit correlation patterns \cite{borges21}.

\subsection{Fractional Fourier transform of photon pairs}

\begin{figure}[h!]
	\centering 	\includegraphics[width=0.7\columnwidth]{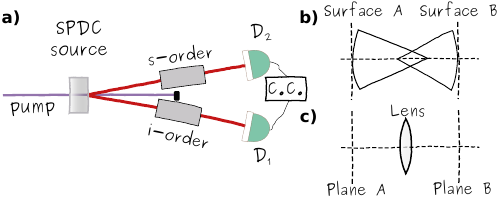}
	\caption{Fractional Fourier Transform of photon pairs: a) illustration of experiments exploring spatial correlations using the FRFT with arbitrary orders for signal (s) and idler (i). Realization of the FRFT in b) Fresnel diffraction and c) using a simple lens system.}
	\label{fig:FRFT}
\end{figure}

A series of experiments performed as part of the PhD thesis of Daniel Tasca, supervised by Steve Walborn and published in Refs. \cite{tasca08,tasca09a,tasca09b}, investigated correlations of photon pairs using the Fractional Fourier Transform (FRFT) \cite{ozaktas01}. This generalization of the Fourier Transform was introduced to us by the visiting French physicist Pierre Pellat-Finet, then at the University of Bretagne Sud in Lorient, France. Around the same time, Fabricio Toscano, then a postdoctoral researcher at UFRJ, became an enthusiast of the FRFT and contributed significantly to the conception and interpretation of several experiments.

The FRFT can be understood in several ways. In phase space, constructed from dimensionless position and momentum variables of a single degree of freedom, it corresponds to a rotation by an angle $\theta$, such that $x \rightarrow x\cos\theta + p\sin\theta$ and $p \rightarrow p\cos\theta - x\sin\theta$. In optical diffraction, the propagation of a field from one spherical surface to another at a specific distance can be described as an FRFT \cite{pellat-finet94}. Similarly, a simple lens system can eliminate the complications of spherical surfaces, allowing one to treat propagation between two transverse planes as an FRFT, where the transform order $\theta$ is determined by the distance between the input and output planes together with the focal length of the lens \cite{ozaktas01}.

The first experiment performed using the FRFT to measure spatial correlations manifested the violation of the Bell-inequality. However, after completion, the authors learned that their experiment was conceptually flawed, and that the violation was actually due to a post-selection bias in the detection system.  Fortunately, this error led the authors to many new ideas concerning the FRFT, one of which being the realization that the flawed experiment was a  quantum simulation of Popescu-Rohrlich correlations \cite{tasca09b}, which are stronger-than-quantum correlations at the forefront of the study of nonlocality \cite{popescu94}.

Previous to the experiments of Tasca \textit{et al.}, experimental observation of spatial entanglement of photon pairs considered only the near and far-field of the non-linear crystal source \cite{howell04,almeida05,almeida06}. Tasca \textit{et al.} used the FRFT to consider arbitrary propagation distances parameterized by $\theta_s$ and $\theta_i$ of the signal and idler photons. They showed that spatial correlations or anticorrelations could be observed when the sum $\theta_s+\theta_i$ was an even or odd multiple of $\pi$, respectively.  Thus, for any propagation parameter $\theta_s$ of the signal photon, a corresponding propagation $\theta_i$ of the idler photon can be found to produce correlation or anti-correlation.  Moreover, pairs of parameters $\theta_s,\ \theta_i$ defining mutually unbiased bases were identified, allowing entanglement to be revealed through the violation of established inequalities \cite{tasca08,tasca09a}. Subsequently, it was demonstrated that entanglement can also be observed using measurements involving three mutually unbiased bases \cite{paul16}. 

\subsection{Beyond Gaussian entanglement and EPR-steering}
\label{sec:beyond}

A series of experiments were carried out at UFRJ as part of the PhD thesis of Rafael Gomes, under the supervision of Paulo H. Souto Ribeiro, focusing on Gaussian and non-Gaussian entanglement in continuous-variable systems. On the theoretical side, Alejo Salles (PhD student supervised by Ruynet Mattos) and Fabricio Toscano collaborated closely with Steve Walborn and the experimental team. Together, they explored the rich structure of non-Gaussian entangled states in the spatial degrees of freedom of photon pairs generated via SPDC. These studies, which formed the core of Rafael Gomes’s PhD work, showed that quantum correlations -- manifesting as entanglement and EPR steering -- can arise in regimes inaccessible to Gaussian-state methods, underscoring the need for more general theoretical tools in continuous-variable quantum optics.

\begin{figure}[h!]
	\centering 	
        \includegraphics[width=0.5\columnwidth]{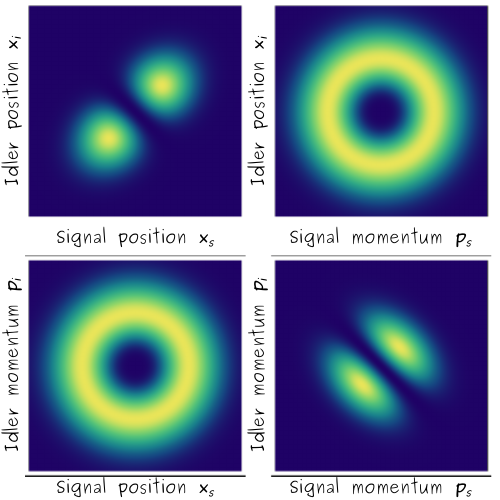}
	\caption{Beyond Gaussian Entanglement and EPR-steering: simulated coincidence count distributions for the two-photon state produced with a HG pump beam. A diagonal HG-like pattern occurs when signal and idler are observed in the same basis, while a LG-like pattern occurs for different bases.}
	\label{fig:HG-state}
\end{figure}

Experimentally, spatial quantum state engineering was achieved through angular spectrum transfer, producing states closely related to those obtained in earlier work with first-order Hermite-Gaussian (HG) pump beams (simulated results are shown in Fig. \ref{fig:HG-state}). To probe genuine non-Gaussian behavior, however, the pump beam was focused onto the nonlinear crystal, thereby suppressing the standard position and momentum correlations (or anticorrelations). In this regime, entanglement originated instead from coherent superpositions of spatial modes. 

Some of the aforementioned researchers reported the experimental generation of a spatially entangled but non-Gaussian two-photon state using standard SPDC, where the non-Gaussianity arose from manipulating the pump beam profile and detection geometry \cite{gomes09}. Rather than relying on Gaussian entanglement criteria -- such as the Duan or Mancini conditions \cite{duan00,mancini02}, which test entanglement via second-order moments -- the authors applied the more general Shchukin-Vogel criteria \cite{shchukin05}, which involve a hierarchy of inequalities based on higher-order moments. Using fourth-order moments via this approach allowed them to detect entanglement in a state where all second-order (Gaussian) criteria fail, thus experimentally demonstrating quantum entanglement that is genuinely non-Gaussian in nature. The work highlighted a fundamental limitation of widely used Gaussian tools and helped establish the importance of higher-order statistical moments in characterizing complex entangled states.

Building on this, a very similar experimental setup was used to study a closely related problem in the context of EPR steering \cite{walborn11}. While the previous study focused on entanglement, this study questioned whether EPR-steering correlations -- originally introduced by Schrödinger and central to one-sided device-independent protocols \cite{branciard12} -- could also be hidden from Gaussian detection techniques. They introduced and experimentally tested a new entropic EPR steering inequality, based on the Shannon entropy of conditional probability distributions in continuous position and momentum measurements. The authors showed that a spatially entangled photon state, which did not violate Gaussian EPR-steering criteria based on conditional variances (such as the Reid criterion \cite{reid89}), did violate the new entropic inequality, thereby revealing steering correlations that were otherwise hidden. This result not only provided a more sensitive and general tool for detecting quantum correlations, but also gave new operational meaning to EPR steering in the non-Gaussian regime, where many real-world systems naturally reside. It was this new entropic inequality that clarified the connection between steering and quantum cryptography, ultimately giving rise to the concept of one-sided device-independent quantum key distribution \cite{branciard12}.

Parallel to these investigations into hidden correlations, the authors explored the topological structure of the same spatially entangled state. Namely, they demonstrated that an optical vortex -- a phase singularity associated with orbital angular momentum (OAM) -- can be encoded nonlocally, across the spatial degrees of freedom of an entangled photon pair \cite{gomes09a}. Neither photon individually exhibited a vortex in its spatial mode, but the joint two-photon wavefunction possessed a phase singularity, which can be considered as a visual manifestation of entanglement. This experiment revealed how nonlocal quantum correlations can give rise to global spatial structures that have no counterpart in classical optics. In Ref. \cite{gomes11}, the authors interpreted these results as the realization of a nonlocal mode converter, converting the HG mode into a LG via an astigmatic lens system acting on the joint two-photon state.

\subsection{Discrete mode entanglement and correlations}

\subsubsection{Vortex thermal states}

Many works have shown that twin photons can perform well as an experimental platform for Quantum Thermodynamics. This happens not directly by heating the photons, but by creating specific thermal states and emulating interactions of them with heat baths, for instance. Ref.~\cite{haffner20} presents a demonstration of remote preparation of Gibbs states analogues of heralded single photons using twin photons from SPDC. The thermal distribution concerns the orbital angular momentum of single photons, so that the zero-order Gaussian mode corresponds to the ground state and excited states correspond to modes with increasing topological charges $\ell\hbar$, where $\ell$ is the orbital angular momentum. Thomas H\"affner and Guilherme Zanin were PhD students at UFSC, under the supervision of Paulo H. Souto Ribeiro. Rafael Gomes and Lucas Céleri were collaborators from  UFG at Goiânia, Brazil, and completed the research team in this work. 

\begin{figure}[h!]
	\centering 	\includegraphics[width=0.7\columnwidth]{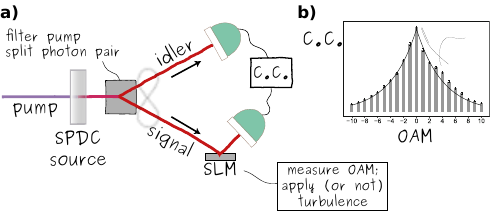}
	\caption{Vortex Thermal states: a) conceptual representation of the remote preparation of thermal states in a SPDC photon and b) experimental results \cite{haffner20}. The signal's OAM distribution was measured by applying raising and lowering OAM masks with the SLM and coupling the resulting beam into a single-mode fiber.}
	\label{fig:thermal}
\end{figure}

In the experiment, measurements performed on the idler photon prepare its twin's (signal photon) state (see Fig.~\ref{fig:thermal}). By choosing to pump the SPDC process with a Gaussian beam and simply detecting the arrival time of the first photon, the authors demonstrate that the orbital angular momentum of the signal photon follows the thermal distribution of a Gibbs state. The associated temperature is controlled by either changing the pump diameter or the detector opening. They could also prepare coherent thermal states in the signal photon by simply projecting the first photon angular momentum before detecting it. The utility of this platform was illustrated by submitting the signal thermal state to an optically simulated turbulence, after which the associated temperature was found to be higher.

\subsubsection{Accelerated entanglement}
\begin{figure}[h!]
    \centering
    \centering 	\includegraphics[width=0.7\columnwidth]{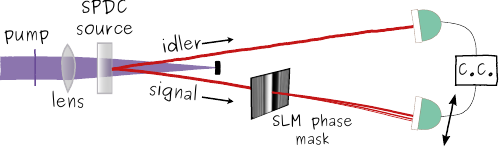}
    \caption{Accelerated Entanglement: Experimental setup with the SLM imposing a cubic phase on the signal beam, creating the Airy pattern.}
    \label{fig:airy}
\end{figure}

Studying the effects of curved spacetime presents formidable experimental challenges. In a creative workaround, Pimenta \textit{et al.} \cite{pimenta24} ~probe these effects by artificially accelerating one photon of an entangled pair and observing its impact on their quantum correlations. The work team counted on Raphael Pimenta as a PhD student and Adriano Barreto as a postdoc at UFSC, under the supervision of Paulo H. Souto Ribeiro. Lucas Céleri from UFG completed the team. The experimental setup used is illustrated in Fig. \ref{fig:airy}. Using SPDC, they generate photon pairs entangled in their transverse spatial degrees of freedom.  A spatial light modulator (SLM) imposes an Airy-beam phase profile \textit{only} on the idler photon, curving its trajectory via diffraction and establishing an optical analog of gravitational acceleration. This setup directly tests whether entanglement survives when one photon experiences non-inertial motion while its partner propagates freely.

Through rigorous theoretical modeling and experiment, the authors confirm the persistence of entanglement. Non-separability is witnessed using the Mancini-Giovannetti-Vitali-Tombesi (MGVT) criterion, $\Delta(p_1 + p_2)\, \Delta(x_1 - x_2) < 1$, where $\Delta(p_1 + p_2)$ is the standard deviation of the sum of the signal's and idler's transverse momenta, and $\Delta(x_1 - x_2)$ is the standard deviation of the difference of the signal's and idler's positions at the detector plane when it is imaged to the crystal plane. Remarkably, despite Airy acceleration distorting the idler's profile, broadening detection statistics and shifting peak positions, the measured variance product consistently violates the separability threshold (ranging from \(0.21\) to \(0.39\) across trials). This confirms that entanglement adapts to the idler's curved path without degradation.

 Accelerating \textit{only} the signal simulates a scenario where one particle experiences pseudo-gravitational acceleration while its partner remains inertial. Airy beams are ideal here, propagating without diffraction spreading to mimic ``rigid'' particles under uniform acceleration. Notably, post-selected coincidence measurements exclusively analyze surviving entangled pairs, inherently filtering photons lost during SLM modulation.

By merging quantum optics with gravitational analogies, this work establishes a practical laboratory platform to explore entanglement in curved spacetime. With direct gravitational tests remaining prohibitively difficult, Airy acceleration offers a compelling proxy. The results suggest entanglement resists certain relativistic distortions, advancing fundamental understanding of quantum-gravity interfaces and informing future technologies like satellite-based quantum networks traversing Earth's gravitational field. 

\section{Entanglement and Decoherence}

During the 1990's, as the experimental techniques in SPDC advanced in Brasil, several Brazilian theoreticians were producing notable work in quantum optics and quantum information.  In the mid-2000's, with the help of a common language provided by quantum information concepts, several quite fruitful collaborations were established between theory and experiment. These studies focused mostly on entanglement and decoherence.

\subsection{Single shot measurement of entanglement}

Entanglement is typically hidden behind layers of quantum state reconstruction and inference. Walborn. et. al. experiment shows it can be determined directly \cite{walborn06}. Their work presents the first direct experimental measurement of concurrence using only a single local measurement, that is, without a full state tomography. Steve Walborn was a postdoctoral at UFRJ, Florian Mintert was also a postdoctoral at the Max Planck Institute in Dresden, Germany, under the supervision of Andreas Buchleitner and he was visiting UFRJ at the time the experiment was proposed together with Luiz Davidovich. Paulo H. Souto Ribeiro and Steve Walborn performed the experiment.

The standard definition of concurrence for pure states involves the nonphysical operation of complex conjugation followed by $\sigma_y \otimes \sigma_y$, which cannot be implemented directly. The authors instead use a two-copy formulation for pure states, where concurrence is given by $
C = 2 \sqrt{P_A}$, where $P_A$ is the probability of projecting the two copies of subsystem A in the antisymmetric state, requiring no measurement on subsystem B.

\begin{figure}[h!]
    \centering
    \centering 	\includegraphics[width=0.7\columnwidth]{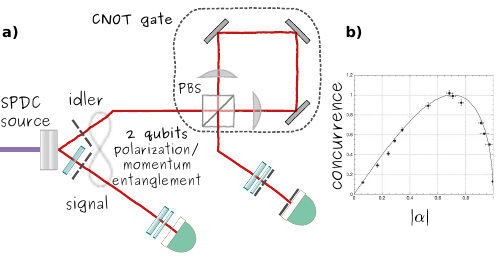}
    \caption{Single shot measurement of entanglement: a) conceptual representation of preparation of two qubits and CNOT-gate-enabled entanglement detection via a single measurement. b) Observed entanglement quantifier for different twin photon states.}
    \label{fig:single-shot}
\end{figure}

The authors encode two identical copies of a quantum state within a pair of photons, using their polarization and transverse momentum degrees of freedom. This enables direct extraction of concurrence by projecting the two qubits (momentum and polarization) of one photon onto the antisymmetric subspace of one subsystem. In the experiment, twin photons were generated via SPDC, such that two simultaneous copies of the state $\ket{\Psi} = \alpha \ket{01} + \beta \ket{10}$ were prepared by tuning the polarization components using wave plates and the momentum components using square apertures (see Fig.~\ref{fig:single-shot}).

To access $P_A$, the authors implemented a polarization-controlled CNOT gate via a polarization-sensitive Sagnac interferometer containing two cylindrical lenses. These induce a relative $180^\circ$ rotation between horizontal and vertical polarizations, transforming the single-photon Bell states into separable combinations of polarization and momentum, which can easily be detected with polarizers and a detector in the momentum modes.
The experimental data closely matched the theoretical expression
$
C = 2|\alpha| \sqrt{1 - |\alpha|^2}
$. 
\par
With this approach, the authors demonstrated that nonlinear entanglement measures, which previously could be considered experimentally inaccessible, can be inferred directly.  While the measurement apparatus itself is linear, the nonlinearity of the entanglement measure is accessed through statistical correlations across the multiple  copies.  
\par
Ideally, one would approach the experimental demonstration of these ideas by producing two pairs of entangled photons simultaneously using a high-power pulsed laser. Since the authors did not have this equipment, they devised the scheme to encode two copies in a single "hyperentangled" photon pair.  In addition, one of the detectors used in the experiment (borrowed from C. H. Monken lab/UFMG) had faulty optics, meaning that it needed to be aligned at an angle, so it appeared to not be ``looking'' at the crystal source. Nonetheless, the authors were able to make it work. This experiment played an important part in boosting the collaboration between experimentalists and theoreticians in quantum information in Brazil.
\par

\section{Dynamics of Entanglement}

\subsection{Entanglement sudden death}

``Entanglement Sudden Death'' (ESD) is a term introduced by T.~Yu and J.~H.~Eberly in their seminal work~\cite{eberly06}. They demonstrated that entanglement decay can be fundamentally different from single-qubit decoherence, showing a strong dependence on the initial state of the composite quantum system. This theoretical insight was experimentally tested in 2007 by a group of theoreticians and experimentalists at UFRJ ~\cite{almeida07}. Marcelo Almeida was a postdoctoral under the supervision of Paulo H. Souto Ribeiro and together with Steve Walborn, Fernando de Melo, Alejo Salles, Malena Hor-Meyll and Luiz Davidovich conceived and implemented the experimental test. They  used twin photons to observe ESD in the laboratory. This experiment was noteworthy in that it was one of the first to use multiple degrees of freedom of photons to emulate open quantum systems, an idea originated, for the most part, in Brazil from discussions between quantum optics experimentalists and theorists such as Luiz Davidovich and Marcelo Fran\c{c}a Santos, at the time at UFMG, and which led to a decade of theoretical/experimental collaboration. The experiment focused on bipartite quantum systems under independent amplitude damping channels, revealing that entanglement can disappear abruptly --- rather than asymptotically --- depending on the initial state's composition.

Specifically, in atomic systems, ESD occurs when the initial entangled state has a greater population in the excited state $|e\rangle$ than in the ground state $|g\rangle$. While individual qubits always decay asymptotically, entanglement vanishes suddenly for states of the form 
$
|\Phi\rangle = \alpha |gg\rangle + \beta e^{i\delta} |ee\rangle
$
when the condition $ |\beta| > |\alpha| $ is met. This behavior highlights the distinctive nature of entanglement dynamics in comparison to single-particle decoherence.

For such states, the concurrence evolves as
$
C = \max\left\{0, 2(1-p)|\beta|\left(|\alpha| - p|\beta|\right)\right\}
$.
Sudden death occurs at the finite value $ p = |\alpha|/|\beta| $, as identified in~\cite{eberly06}, when the term $ |\alpha| - p|\beta| $ becomes negative. Physically, this reflects the greater vulnerability of entanglement to sudden collapse when the initial state favors the double-excitation component.

\begin{figure}[h!]
    \centering
    \centering 	\includegraphics[width=0.7\columnwidth]{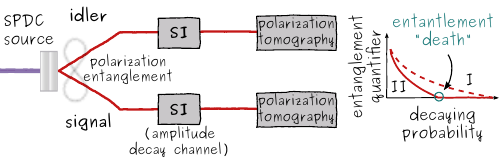}
    \caption{Entanglement Sudden Death: conceptual representation of the experimental setup used to study entanglement dynamics when each photon is submitted to an amplitude-damping reservoir. SI: Sagnac Interferometer.}
    \label{fig:sudden-death}
\end{figure}

In the experiment by Almeida \textit{et al.}~\cite{almeida07}, a pair of polarization-entangled photons generated by SPDC were used to simulate the entangled atomic state $|\Phi\rangle$. Each photon encodes a two-level system --- mapping H and V polarization states onto the ground and excited states, respectively --- while a Sagnac interferometer (SI) with two distinguishable optical paths simulates the interaction of the two-level system with a reservoir. The complex amplitudes $\alpha$ and $\beta$ are prepared as polarization amplitudes when they produce the polarization-entangled biphoton and are varied with waveplates and source alignment.
The critical dependence on the asymmetry of the initial state was confirmed using two entangled states with identical initial concurrence but different amplitude ratios (see Fig.~\ref{fig:sudden-death}). For \textbf{State I} (\( |\beta|^2 = |\alpha|^2/3 \)), where the ground state dominates (\( |\beta| < |\alpha| \)), entanglement decayed gradually and vanished only at \( p = 1 \), when both qubits had fully decayed. In contrast, \textbf{State II} (\( |\beta|^2 = 3|\alpha|^2 \)) exhibited sudden death at \( p \approx 0.55 \), close to the predicted threshold \( p = 1/\sqrt{3} \), due to its excited-state dominance (\( |\beta| > |\alpha| \)). These results demonstrate that the relative population of excited states in the initial superposition determines whether entanglement decays abruptly or smoothly.

The underlying mechanism lies in how amplitude damping affects entangled superpositions. When \( |\beta| > |\alpha| \), decay events quickly disrupt the fragile coherence between \( |gg\rangle \) and \( |ee\rangle \), populating intermediate separable states such as \( |eg\rangle \) and \( |ge\rangle \). A larger initial excited-state component increases the likelihood of this disentangling pathway. Notably, this effect is specific to dissipative (amplitude damping) environments. Control experiments with dephasing channels showed that in such cases, entanglement always decayed asymptotically, regardless of the initial state. Thus, the study conclusively shows that ESD is not universal, but arises from the interplay between specific initial conditions and the nature of the decoherence process.  From a technical standpoint, the experiment by Almeida \textit{et al.}~\cite{almeida07} was perhaps the first to use the double-path Sagnac interferometer in a Quantum Optics/Information setting, a setup that has been used by many groups since.
\par
In other studies, also conducted in Brazil, similar phenomena have been observed in continuous-variable optical systems~\cite{Barbosa2010}, and related approaches have been used to study decoherence in photonic qudits~\cite{Marques2015}.

 \subsection{Emergence of the pointer basis}

At the beginning of the 2010s, the quantum-to-classical transition and the role of decoherence in quantum measurement were already well-established topics in the foundational literature. Seminal theoretical contributions had introduced the concept of the pointer basis -- preferred states of a quantum system that remain stable under environmental interaction -- and demonstrated how decoherence tends to suppress coherences in a specific basis, making quantum systems appear classical to an observer. However, most existing analyses focused on asymptotic regimes, where the system had already lost all quantum coherence. Experimental studies, while numerous, often centered on entanglement decay or loss of visibility, without clearly connecting these effects to the emergence of classical correlations in measurement contexts. A formal, experimentally accessible notion of the dynamical emergence of the pointer basis -- especially in a non-asymptotic regime -- was still lacking.

In 2012, a fruitful cooperation between the theoretical group of UNICAMP led by Profs. Marcos de Oliveira and Amir Caldeira, and postdocs Marcio Cornelio and Felipe Fanchini and the experimental team of UFRJ led by Profs. Steve Walborn and Paulo H. Souto Ribeiro with postdocs Osvaldo Jimenez-Farias, Malena Hor-Meyll and Irenée Frerot addressed this gap by offering both a theoretical framework and an experimental demonstration of how classical correlations between a quantum system and its measurement apparatus evolve under decoherence, with the paper ``Emergence of the Pointer Basis through the Dynamics of Correlations'' \cite{cornelio12}. The key insight is that these correlations can undergo a sudden transition from a regime of decay to a regime where they plateau. This behavior signals the moment when the pointer basis emerges: a set of apparatus states that remain robust and reliably correlated with the system's states. Importantly, this emergence occurs while the system–apparatus composite can still retain quantum correlations, indicating that classicality appears locally (in the apparatus) before global decoherence is complete.

\begin{figure}[h!]
	\centering 	\includegraphics[width=0.7\columnwidth]{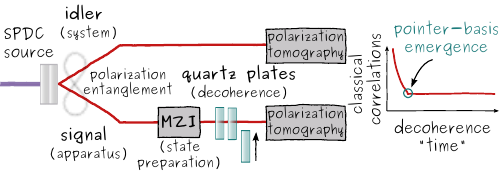}
	\caption{Emergence of the Pointer Basis: conceptual representation of the experimental setup used to show the emergence of a pointer basis during the dynamics of decoherence, emulated by the introduction of quartz plates. MZI: Mach-Zehnder Interferometer.}
	\label{fig:pointer-basis}
\end{figure}

To test this prediction, the authors implemented an SPDC source that generated polarization-entangled photons, as depicted in Fig.~\ref{fig:pointer-basis}. The idler photon represented the quantum system, and the signal photon functioned as the measurement apparatus. A crucial step in the experiment is the preparation of an incoherent mixture of the four Bell states, achieved by introducing a Mach-Zehnder interferometer in the path of the signal photon ($=$ apparatus). This setup scrambles the relative phases among the entangled components, simulating a measurement-like scenario in which coherence between pointer states is lost.

To introduce controlled decoherence, the authors implemented a phase damping channel by passing the signal photon through quartz crystal plates positioned after the interferometer. These birefringent elements induce polarization-dependent delays, mimicking the environmental interaction with the apparatus. The total thickness of the quartz corresponds to an effective time parameter, enabling the study of dynamical evolution. By analyzing the classical correlations between the system and the apparatus as a function of this effective time, the experiment revealed a distinct transition point, corresponding to the emergence of the pointer basis. This result provided a novel and concrete signature of classicality appearing in a quantum system and deepened our understanding of the interplay between information, decoherence, and measurement.

\subsection{Non-Markovianity}

The study of non-Markovian dynamics in open quantum systems gained significant attention beginning around 2009, when researchers introduced formal criteria to characterize memory effects in quantum evolutions. Unlike traditional Markovian processes, where information irreversibly leaks into the environment, non-Markovian behavior allows for partial information to flow back into the system, reviving quantum coherence and entanglement. Breuer \textit{et al.}, proposed a widely adopted measure based on trace distance, linking non-Markovianity to the temporary increase in distinguishability between quantum states \cite{breuer2009}. Shortly after, Rivas \textit{et al.} introduced a complementary criterion, associating non-Markovianity with the breakdown of completely positive divisibility and the revival of entanglement with an ancillary system \cite{rivas2010}.

\begin{figure}[h!]
        \centering
	\centering 	\includegraphics[width=0.7\columnwidth]{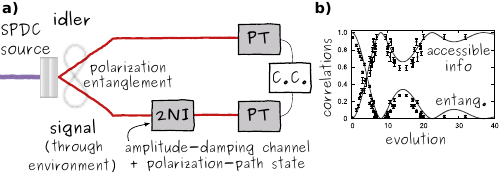}
	\caption{Non-Markovianity: a) conceptual representation of the experimental setup used to detect the flow of information between twin photons and an environment. 2NI: two nested interferometers; PT: polarization tomoghraphy.  b) Accessible information and entanglement revivals obtained in \cite{fanchini14}.}
	\label{fig:non-markovianity}
\end{figure}

Non-Markovianity with entangled photons was explored by the collaboration involving Marcos de Oliveira from UNICAMP, Felipe Fanchin from UNESP,  G. Karpat and B. Çakmak from Sabanci University in Turkey, Leonardo Castellano from UFSCar, and Gabriel Aguilar, Osvaldo Jimenez-Farias, Steve Walborn and Paulo H. Souto Ribeiro from UFRJ. Their joint work introduced a novel entanglement-based measure using accessible information ($J_{\overleftarrow{\mathcal{SE}}}$), derived from the classical knowledge of the environment about a system $\mathcal{S}$ via an apparatus $\mathcal{A}$. The experiment, illustrated in Fig.~\ref{fig:non-markovianity}, employed polarization-entangled photons in two nested interferometers, where beam displacers simulated an amplitude damping channel for one photon, while tomography tracked tripartite states across the system $\mathcal{S}$, $\mathcal{A}$ (damped photon), and $\mathcal{E}$ (path environment). The key results showed that $J_{\overleftarrow{\mathcal{SE}}}$ monotonically increased in the Markovian regimes but decreased during non-Markovian revivals, directly linking the revival of entanglement ($E_{\mathcal{SA}}$) to the flow of information.

In Ref. \cite{fanchini14}, published in the Physical Review Letters, the authors presented an optical implementation, which provides full access to the environment, validated the Koashi-Winter relation \cite{koashi2004}($ E_{\mathcal{SA}} = S(\rho_{S}) - J_{\overleftarrow{\mathcal{SE}}} $) and established $J_{\overleftarrow{\mathcal{SE}}}$ as an entropic witness of the information flow. 

\section{Conclusion}

In conclusion, we have traced the rise of twin-photon experiments in Brazil, beginning with the pioneering work of Prof. Geraldo A. Barbosa at UFMG in Minas Gerais during the early 1990s. We have highlighted the remarkable impact and advances that this line of research has brought to Quantum Optics and Quantum Information, both in Brazil and worldwide. The selected experiments described here reflect the historical evolution of this approach in Brazil, as seen from the perspective of the authors.

Although our focus has been entirely on twin-photon experiments, it is important to note that other closely related systems and experimental techniques have also undergone significant development and impact in Brazil and abroad. Among them are twin beams produced with Optical Parametric Oscillators (OPO) \cite{Martinelli2004,Villar2005}, the generation of entangled photon pairs via Raman scattering \cite{Saraiva2017}, and the production of quantum-correlated light beams in atomic vapors \cite{Marinho2025,Araujo2022}.

In this International Year of Quantum Science and Technology (IYQ), we emphasize the relevance and breadth of the contributions of Brazilian Twin Photonics to the development of a strong scientific critical mass in Brazil, one that is now actively working to drive the transition from fundamental science to practical applications.

\section*{Acknowledgements}
This work was supported by the following Brazilian research agencies: Conselho Nacional de Desenvolvimento Científico e Tecnológico (CNPq - DOI 501100003593), Coordenação de Aperfeiçoamento de Pessoal de Nível Superior (CAPES DOI 501100002322), Fundação de Amparo à Pesquisa do Estado de Santa Catarina (FAPESC - DOI 501100005667), Instituto Nacional de Ciência e Tecnologia de Informação Quântica (INCT-IQ, CNPq Grant No. 465469/2014-0), Instituto Nacional de Ciência e Tecnologia de Infraestruturas Quântica e Nano para Aplicações Convergentes (INCT-IQNano, CNPq Grant No. 406636/2022-2), and Instituto Nacional de Ciência e Tecnologia em Dispositivos Quânticos (INCT-QD, CNPq Grant No. 408783/2024-9).  SPW and GHS were supported by Fondo Nacional de Desarrollo Científico y Tecnológico (FONDECYT) Grant No. 1240746 and ANID -- Millennium Science Initiative Program -- ICN17$_-$012.

\bibliography{bibfile}

\end{document}